\documentclass[secnumarabic,showpacs,amssymb, amsmath, nofootinbib,
nobibnotes, aps,article]{revtex4}

\usepackage{amsmath,amssymb}
\usepackage{graphicx}

\begin{document}

\title{Partial rejuvenation of a colloidal glass}
\author{F. Ozon$^1$}
\author{T. Narita$^1$}
\author{A. Knaebel$^1$}
\author{G. Debr{\'e}geas$^2$}
\author{P. H{\'e}braud$^3$}
\author{J.-P. Munch$^1$}

\affiliation{
$^1$ L.D.F.C, UMR $7506$, $3$ rue de l'Universit{\'e}, $67084$ Strasbourg Cedex, France\\
$^2$ L.F.O., UMR $7125$, Coll\`ege de France, $11$ place Marcelin Berthelot, $75231$ Paris cedex $05$ France\\
$^3$ L.P.M. ESPCI, UMR $7615$, $10$ rue Vauquelin, $75231$ Parix Cedex $05$, France\\
}

\begin{abstract}
We study the effect of shear on the aging dynamics of a colloidal suspension of synthetic clay particles.
We find that a shear of amplitude $\gamma$ reduces the relaxation time measured just after the cessation of shear
 by a factor $\exp(-\gamma/\gamma_c)$, with $\gamma_c \sim 5\%$, and is independent of the duration and
 the frequency of the shear. This simple law for the rejuvenation effect shows that the energy involved
 in colloidal rearrangements is proportional to the shear amplitude,
 $\gamma$, rather than $\gamma^2$, leading to an Eyring-like
 description of the dynamics of our system.
\end{abstract}

\maketitle

\section{Introduction}

Molecular and spin glasses exhibit very intriguing dynamical
properties, which are still poorly understood. One of the most
striking features is the aging phenomenon~: a slowing down of the
microscopic dynamics with the elapsed time. This behavior is
generally associated with the existence of many accessible
metastable states spanning a broad energy distribution. As time
goes on, the material gets trapped in deeper and deeper energy
wells for increasing escaping times.  As a result, many physical
characteristics of these systems (such as
rheological~\cite{struik}, or dielectric properties) depend in a
non trivial way on their thermal history.

In recent years, aging dynamics has been evidenced in very
different systems, such as colloidal
suspensions~\cite{derecmanip,cloitremicrogel}, dense emulsions,
amorphous polymers or weak gels. In these so-called soft glassy
materials, temperature can not be varied without drastically
modifying the underlying weak interactions between their
constituents. It is thus not a convenient parameter to probe such
history dependent effects.  However, it has been recently
suggested that shearing could play an analogous role as
temperature by allowing a renewal of the microscopic
structure~\cite{lequeuxviasnoff, berthierbarrat,liunature}.
Quenching of colloidal glasses for instance, can be obtained by
applying a temporary strong shearing to the material and then
stopping the shear. This temperature/strain analogy has been
extensively developed in different models, and allowed to account
for some rheological features of soft glassy materials.

The relative success of this approach actually hides important
conceptual issues~: What is the effect of a macroscopic strain on
microscopic scales in such disordered systems? Can this mechanical
energy input yield a spatially homogeneous energy term equivalent
to a temperature raise? If so, how does this local energy depend
 on the amplitude, frequency or duration of the shear?

In the present note, we address these questions by measuring slow
relaxation times in a colloidal glass submitted to shear sequences
of various amplitude, frequency and duration. We focus on the
modification of the aging dynamics of the quiescent material after
a finite shear period has been applied to identify the relevant
mechanical parameters.

\section{System and technique}

\begin{figure}[ht]
\includegraphics[height=6cm]{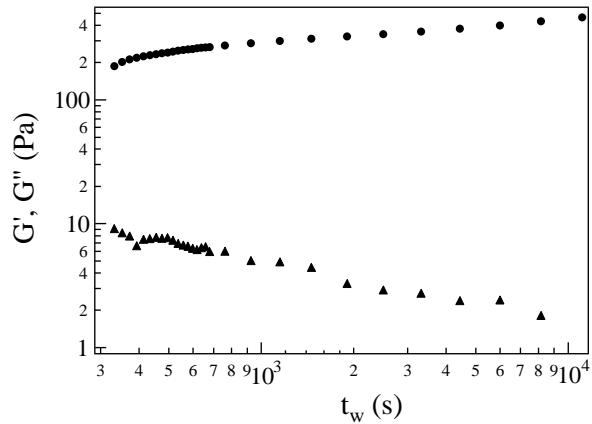}
\caption{Elastic $(\bullet)$ and loss $(\blacktriangle)$ moduli of
a suspension of smectite at $\phi=3\ \%$. }\label{fig:GpGs}
\end{figure}

The system used is a suspension of synthetic smectite particles
(Sumecton SA, kindly provided by Kunimine Industries Co., Ltd.,
Tokyo, Japan) dispersed in water. These clay particles consist in
negatively charged disks of diameter $125\ nm$ and thickness $5\
nm$. The solution is prepared by dispersing the powder in water,
and stirring at room temperature for three days. At $pH=9$ and
concentrations higher than $\phi=1.2\ \%$, they form a stable
solution with a finite elastic modulus that slowly evolves with
time. In all the experiments, the concentration is set to $\phi=3\
\%$ for which we measure an elastic modulus $G'=300\
Pa$~(\textsc{Fig}.\ref{fig:GpGs}). In order to probe the internal
dynamics, we add latex polystyrene particles of diameter $1\ \mu
m$ at a volume fraction of $1\ \%$. These particles render the
system turbid thus allowing for subsequent multiple light
diffusion studies.  The sample is placed in a $1\ mm$ thick
transparent shear cell controlled by a linear stepper motor, and
illuminated with an enlarged laser beam. Dynamics of the particles
is monitored by diffusing wave spectroscopy (MSDWS)\cite{msdws}.
Using a CCD camera, we record the multispeckle pattern at the
outside surface of our sample. The correlation functions of
transmitted intensity were then computed by averaging the
intensity fluctuations over the entire pattern, that consists of
over $5000$ speckles. Though, rather than computing the
correlation function of diffused intensity by averaging over
temporal fluctuations, one averages the intensity correlation over
spatial fluctuations of the speckle pattern. We thus access :
$g_{2t_w}(t)=\frac{\textstyle\langle
I(t_w)I(t_w+t)\rangle}{\textstyle\langle I(t_w)\rangle\langle
I(t_w+t)\rangle}$, where $\langle ...\rangle$ denotes averaging
over the speckle pattern and $t_w$ is the reference time of the
first image and $t$ the time elapsed since $t_w$. Then the field
correlation function at time $t_w$,
$g_{1t_w}(t)=\sqrt{g_{2t_w}(t)-1}$ is computed. This technique
allows to study slow transient phenomena with characteristic times
between $1$ and $5000\ s$, each of them being uncorrelated and $q$-independent.\\

\begin{figure}[ht!]
\includegraphics[height=6cm]{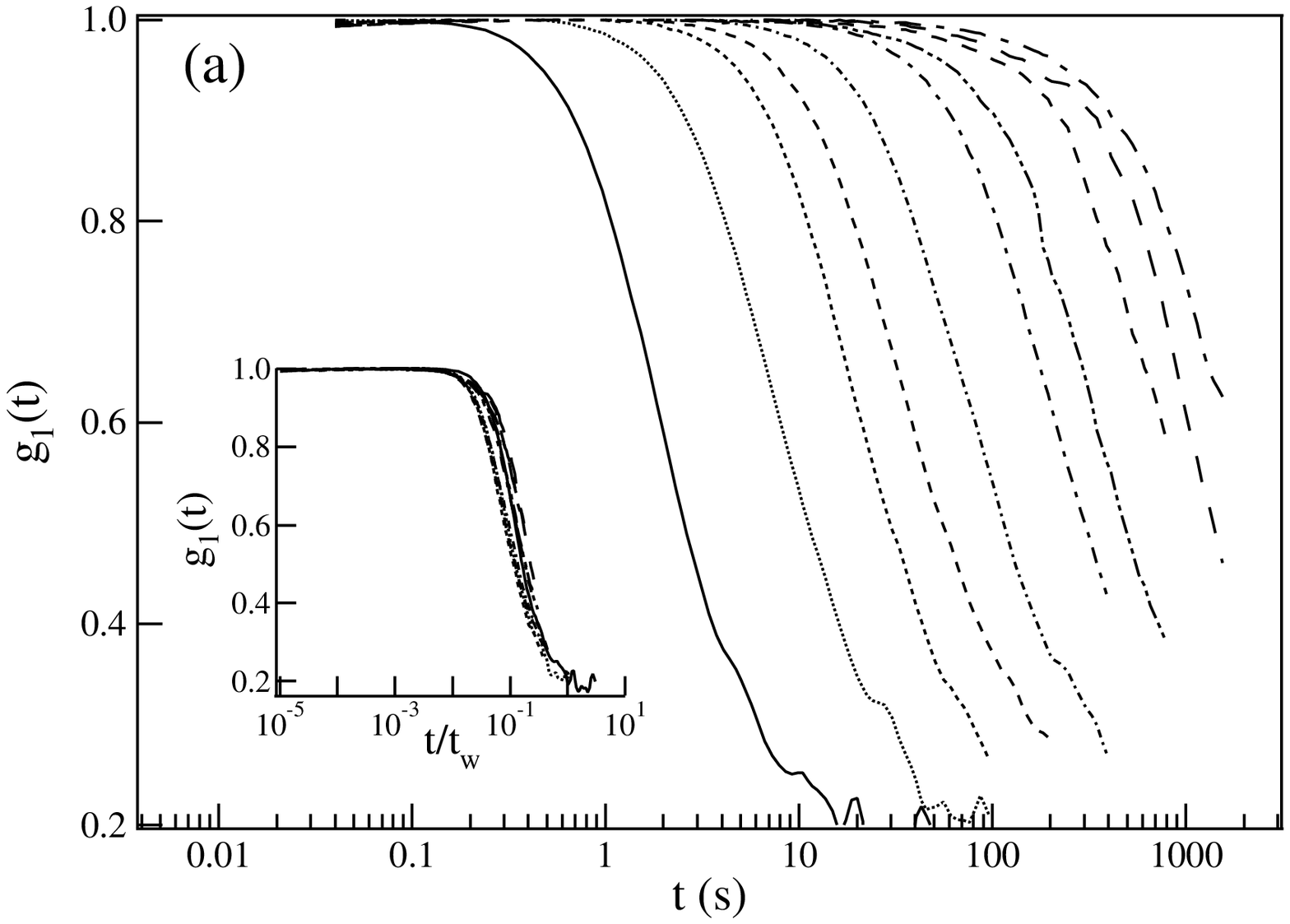}
\includegraphics[height=6cm]{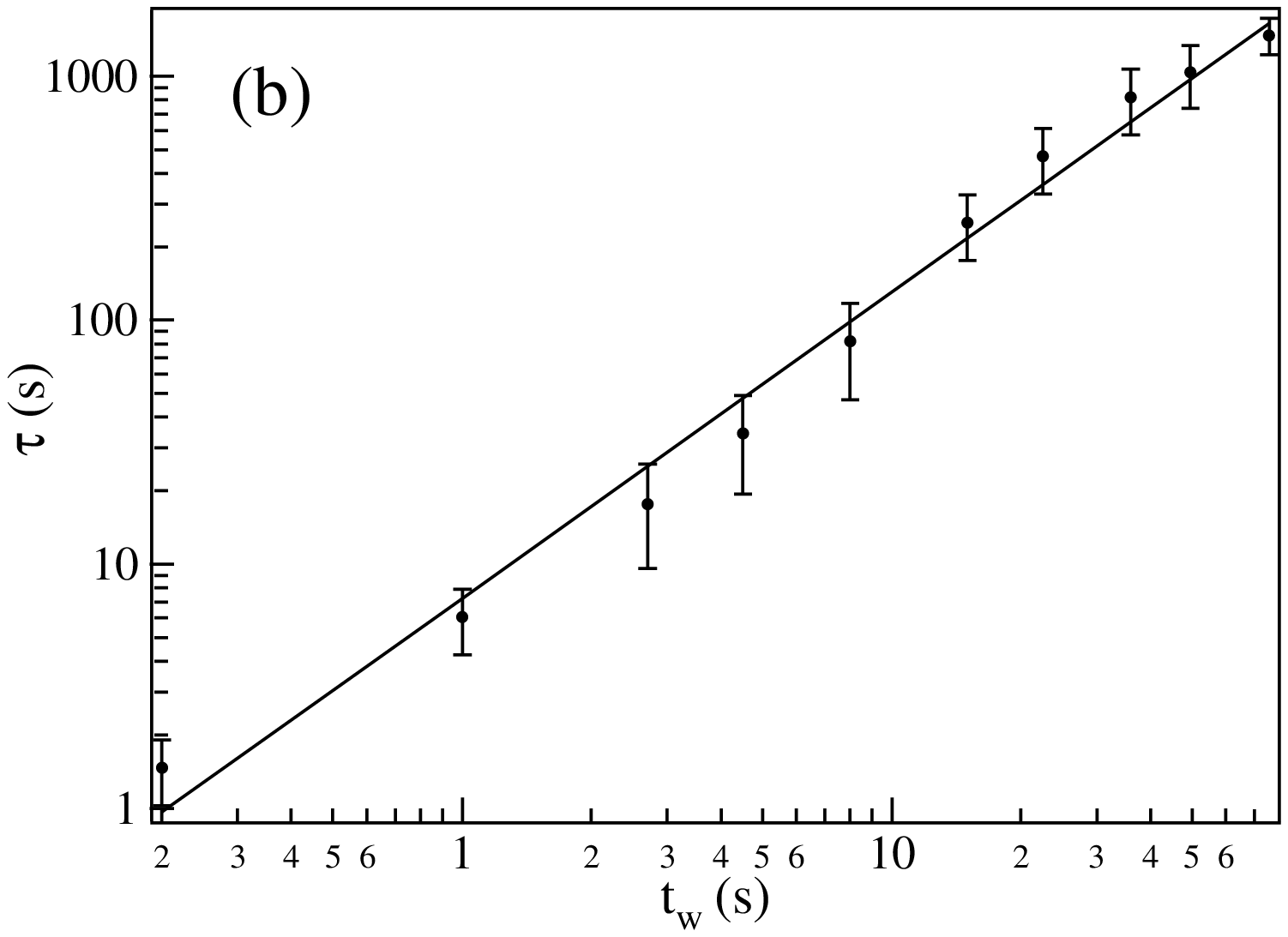}
\caption{(a) Intensity correlation functions at increasing waiting
times, from left to right : $t_w=12s$, $60s$, $162s$, $270s$,
$480s$, $900s$, $1350s$, $2160s$, $2970s$ and $4536s$. {\it
Inset~: }correlation functions {\it vs} rescaled time $t/t_w$. (b)
Decay time $\tau$ of the intensity correlation function as a
function of time spent in the glassy phase, $t_w$. The solid line
is a power-law fit : $\tau\propto t_w^{1.1}$.}
\label{fig:tauaging}
\end{figure}

 In a first series of experiments, we investigate the aging dynamics of the system at rest.
 We prepare the system by submitting it to a sinusoidal
shear strain of amplitude $\gamma=300\ \%$, at frequency $0.3\
Hz$, during $210\ s$. We observed that the turbidity of the system
did not change when shearing was applied. The shear is stopped at
time $t=0$; we then measure the correlation functions of the
intensity at various waiting times $t_w>0$
(\textsc{Fig.}~\ref{fig:tauaging}a). From the slope of these
curves when $t\rightarrow 0$, we extract a correlation time
$\tau$, that increases with time $t_w$. This time $\tau$
corresponds to the slow relaxation processes of glassy dynamics,
called $\alpha$--relaxation. Using a photomultiplier and a fast
correlator, we checked that the fast ($\beta$) relaxation time,
associated with the rapid thermal motion of the particles inside
transient traps, did not depend on the age of the system, as
already observed in a similar system~\cite{knaebel}. We notice
that the plateau value of the correlation function, between fast
and slow decay, is almost constant, which is consistent with our
observation that the elastic modulus of the sample barely evolves
with time.
 We observe that after this strong shearing the relaxation time $\tau$ increases as $t_w^{\alpha}$ with
$\alpha=1.1\pm 0.16$ (\textsc{Fig.}~\ref{fig:tauaging}b). This
value $\alpha \sim 1$ corresponds to a full aging behavior which
has been commonly reported in dense solid suspensions and
non-colloidal glassy systems~\cite{knaebel}. By performing
successive shearing of amplitude $\gamma=300\ \%$ at frequency
$0.3\ Hz$, followed by aging measurements, we obtain reproducible
aging curves, which indicates that the applied shear allows for a
complete rejuvenation of the system. In the rest of this paper, we
focus on the evolution of the slow relaxation time $\tau$ after
shear of moderate amplitudes.

\section{Shear of moderate amplitude}

\begin{figure}[ht!]
\includegraphics*[height=6cm]{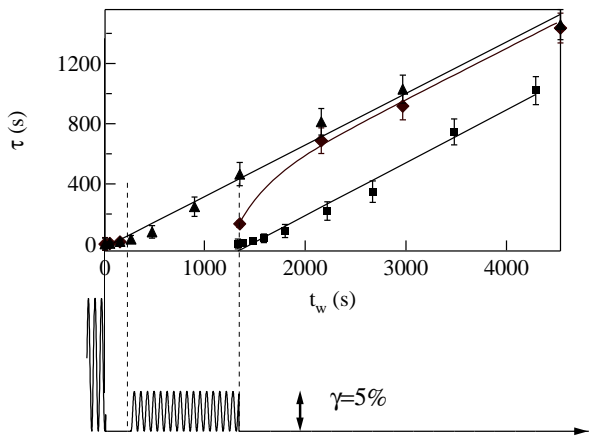}
\caption{Relaxation time after several histories, each of them
being preceded by a phase of full rejuvenation ($300\ \% $ shear
during $210s$). Aging only (no applied shear, $\blacktriangle$).
Aging during $4\ min$, and application of a moderate shear of
amplitude $5\ \%$ during $18\ min$ ($\blacklozenge$). Aging during
$4\ min$, and application of a shear of amplitude  $300\ \%$
during $18\ min$ ($\blacksquare$). The reference time, $t_w=0$ is
taken after the phase of full rejuvenation. Lines are guides to
the eye.} \label{fig:partialaging}
\end{figure}

We thus now apply the following protocol~: after complete
rejuvenation of the sample ($300\ \%$ shear at frequency $0.3\ Hz$
for $180\ s$), we let the system age at rest for $240\ s$ then
apply a moderate
sinusoidal shear.\\
\indent  Let us first study the dynamics of the system after a
moderate shear, of amplitude $\gamma=5\ \% $, frequency $1\ Hz$
and duration $1080\ s$. Rather than following the standard aging
curve for a quiescent material, the dynamics rapidly recover the
original one, corresponding to $\gamma=0$
(\textsc{Fig.}~\ref{fig:partialaging}). $\tau$ is thus not a
sufficient parameter to measure the age of the system, since two
systems with the same relaxation time can evolve differently
depending on the history of their preparation. We thus recover the
same property of the dynamical properties of concentrated
suspensions of colloidal particles in the glassy phase studied in
\cite{lequeuxviasnoff}. It is claimed by these authors that these
different evolutions are related to a modification of the energy
traps {\it distributions} induced by the intermediate shearing.
Identical effects can be observed in spin glasses following a
quench inside the glassy phase.\\

\begin{figure}[ht!]
\includegraphics[height=5.5cm]{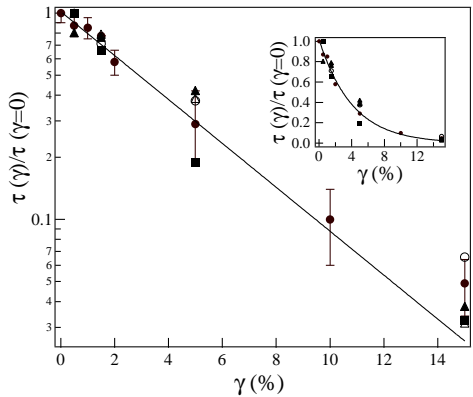}
\caption{Decay time of the intensity correlation function, $\tau$,
measured $60\ s$ after cessation of sinusoidal shear of amplitude
$\gamma$, frequency $\nu$, and duration $t_s$ {\it vs} the shear
amplitude $\gamma$. The absolute value of $\tau(\gamma=0,t_s)$
should be read on \textsc{Fig.}\ref{fig:tauaging}b, at time
$t_w=300+t_s$. $\bullet$~: $\nu=1\ Hz$, $t_s=10\ s$,
$\blacksquare$~: $\nu=.3\ Hz$, $t_s=10\ s$, $\blacktriangle$~:
$\nu=3\ Hz$, $t_s=10\ s$, $\circ$~: $\nu=1\ Hz$, $t_s=1\ s$,
$\vartriangle$~: $\nu=1\ Hz$, $t_s=100\ s$.
  {\it Inset} : linear plot of the data.}
\label{fig:taugamma}
\end{figure}

Dependence of $\tau$ just after cessation of shear, with the
duration of the shear, $t_s$, its frequency, $\nu$ and its
amplitude $\gamma$ is now studied. With our experimental device,
we could vary $\gamma$ between $0$ and $500\ \%$, and $\nu$
between $.3$ and $3\ Hz$. Shear duration $t_s$ was varied between
$1$ and $100\ s$.
  The relaxation time  $\tau$ measured $60\ s$ after shear cessation is found to be decreased relatively
to the value it would have reached if no shear had been applied
(\textsc{Fig.}~\ref{fig:taugamma}). The reduction factor appears
to be a sole function of the amplitude $\gamma$. As shown in
\textsc{Fig.}~\ref{fig:taugamma}, $\tau$ can be written as :

\begin{equation}
\label{tau_of_gamma} \tau(\gamma)/\tau(\gamma=0) =
\exp({-\gamma/\gamma_{c}})
\end{equation}

\noindent with $\gamma_c=.05$ being the typical strain amplitude
necessary to rejuvenate the system.

This result provides a direct quantitative evaluation of the
effect of a moderate shear period on the relaxation dynamics in
soft glassy systems. In particular, we note that  $\tau$ is
independent of the frequency $\nu$  of the applied shear (in the
range $.3$ to $3\ Hz$ explored) and of the shear duration $t_s$
(in the range $1\ s$, $100\ s$). This rules out any cumulative or
frequency dependent effect (at least within our experimental
range) in the way shearing may affect the internal dynamics.

In order to understand the peculiar form of equation
(\ref{tau_of_gamma}), we use a simple physical picture of the
aging process~\cite{bouchaudcomtetmonthus,bouchaudaging}. In this
scheme, the dynamics results from the simultaneous relaxation of
frozen structures of different characteristic energies. We do not
discuss here the microscopic details of these structural
relaxations. They might correspond to different length scales
(cooperative relaxation), or just be associated with different
local yield energy barriers due to the disordered nature of the
suspension. After a time $t_w$ following a quench, we can define a
limiting energy $E(t_w)$ which separates structures of energy
$E<E(t_w)$ that have been fully relaxed, from out-of-equilibrium
structures of energy $E>E(t_w)$. At a given waiting time $t_w$,
the relaxation time $\tau(t_w)$ is controlled by the dynamics of
the yet unrelaxed structures of lowest energy barriers, through a
simple activation type equation : $\tau(t_w)\sim\exp(-E(t_w)/kT)$.

The applied intermediate shearing provides an extra energy input
$\Delta E(\gamma)$ to the system. When the system is quenched
again, structures of energy lower than $\Delta E(\gamma)$ - which
have been entirely rejuvenated - rapidly relax (for $\Delta
E(\gamma)<<kT$, this dynamics rapidly freezes out, so that this
initial period is not accessible to our measurement). Moreover,
the height of the energy barriers greater than $\Delta E(\gamma)$
have been randomly redistributed around their values by a quantity
$\Delta E(\gamma)$. As a result, the apparent energy barrier of
the first accessible relaxation processes becomes $E=E(t_w)-\Delta
E(\gamma)$ and the associated relaxation time has been reduced by
a factor $\exp(-\Delta E(\gamma))$.

Equation (\ref{tau_of_gamma}) demonstrates that the mechanical
energy input $\Delta E(\gamma)$ is proportional to the shear
amplitude $\gamma$. This result is at odds with the $\gamma^2$
dependence used in previous models which assumed a local elastic
response of the material\cite{sgm,sgmlong}.  The correct scheme is
rather similar to Eyring's activated process description of fluid
flows. According to this description, one defines a fluid element
of size $a$, the characteristic interparticular distance. Then,
the fluid flows as soon as this element is deformed by $100\ \%$,
meaning that the energy input to this element, when submitted to
the stress $\sigma$, is $E\sim (\sigma.a^2) a$, {\it i.e.}, if
$G'$ is the elastic modulus of the element, $E\sim G'\gamma
a^3$~\cite{eyring}. We can now estimate the equivalent temperature
increase $\Delta T(\gamma)$ induced by the shear strain of
amplitude $\gamma_c$ that rejuvenates the system. We first
estimate the activation volume as $a^3$ where $a\sim 100\ nm$ is
the typical distance between neighboring particles. The mechanical
energy input then writes $\Delta E(\gamma) = k_B \Delta T(\gamma)
\sim G'\gamma a^3$ where $G'$ is the elastic shear modulus of the
material. Since $\gamma_c=.05$ is the typical strain necessary to
rejuvenate the system, we can in particular estimate that the
deviation from the glass transition necessary to rejuvenate the
system is of the order $T_g-T=G'\gamma_c a^3/k_B \approx T/10$.\\

\section{Conclusion}

We have shown than an externally applied deformation plays the
same role in our colloidal glass as a temperature raise in glassy
liquids, by allowing a restarting of the internal dynamics. We
have demonstrated that the maximum amplitude of the deformation is
the only parameter which controls the magnitude of the
rejuvenation effect~: a very simple correspondence between
temperature and shear amplitude can therefore be drawn.
\\

\footnotesize{We would like to thank F. Thalman for fruitful
discussions. The authors also would like to thank Kunimine
Industries Co. Ltd. for providing the smectite sample.}

\newpage

\newpage

\end{document}